\documentclass[prl,reprint,showpacs,amsmath,amssymb,superscriptaddress]{revtex4-1}
\usepackage{bm}
\usepackage[pdftex]{graphicx}

\usepackage{color}

\begin{document}
\title{Two-Way Coupled Dynamics of Quantized Vortex Bundle Formation in Spatially Periodic Normal-Fluid Turbulence}
\author{Hiromichi Kobayashi}
\affiliation{Department of Physics \& Research and Education Center for Natural Sciences, Hiyoshi Campus, Keio University, 4-1-1 Hiyoshi, Kohoku-ku, Yokohama 223-8521, Japan}
\author{Makoto Tsubota}
\affiliation{Department of Physics, Nambu Yoichiro Institute of Theoretical and Experimental Physics, Osaka Metropolitan University, 3-3-138 Sugimoto, Sumiyoshi-ku, Osaka 558-8585, Japan}
\author{Rio Yokota}
\affiliation{Global Scientific Information and Computing Center, Institute of Science Tokyo, Meguro, Tokyo 152-8550, Japan}
\date{\today}

\begin{abstract}
Using fully coupled two-way simulations combining the vortex filament model with the Navier--Stokes equations, we discover that quantized vortices in superfluid $^4$He are systematically rolled up into coherent bundles within normal-fluid vortex tubes.
Although co-rotating quantized vortices do not aggregate spontaneously, mutual friction with the normal fluid actively drives their bundling, directly inheriting the spatial hierarchy of classical turbulence.
Mutual friction excites high-wavenumber normal-fluid vortex tubes adjacent to quantized vortices, while the superfluid forms a nested structure of thick low-wavenumber outer tubes and fine high-wavenumber inner tubes around the vortex lines.
A distinct time lag between superfluid kinetic energy and dissipation fuels the energy cascade, giving rise to Kolmogorov's $k^{-5/3}$ spectrum.
Crucially, bundle lifetimes are governed by the ratio of the local coupling time scale $t_\mathrm{cl}$ to the Kolmogorov dissipation time scale $\tau_\eta$ ($t_\mathrm{cl}/\tau_\eta < 20$), leading to enhanced thermal stability at higher temperatures.
These findings establish a unified mechanism for two-fluid multiscale vortex organization.
\end{abstract}
\maketitle

{\it Introduction.}---
Quantum turbulence provides a unique platform for exploring the interplay of quantum mechanics and classical hydrodynamics \cite{tsubota25,barenghi23}. Below the superfluid transition temperature, liquid $^4$He is described by the two-fluid model \cite{tisza38,Landau41}, consisting of an inviscid superfluid and a viscous normal fluid. Although this framework successfully explains numerous macroscopic phenomena, understanding turbulence in coupled superfluid and normal-fluid systems remains one of the central challenges in quantum hydrodynamics. In thermal counterflow, the two fluids interact through mutual friction mediated by quantized vortices, giving rise to a rich variety of flow states and transitions \cite{feynman55,vinen57a,vinen57b,vinen57c,vinen58,tough82}.

Earlier theoretical studies based on the vortex filament model (VFM) mainly adopted a one-way coupling approach in which the normal-fluid velocity field was prescribed \cite{schwarz85,schwarz88,adachi10}. These simulations successfully reproduced many characteristics of the TI state, where quantum turbulence develops while the normal fluid remains laminar. However, recent visualization experiments using metastable helium molecules have revealed that increasing the heat flux substantially modifies the normal-fluid flow, leading from Poiseuille flow to a tail-flattened velocity profile and eventually to turbulence \cite{marakov15}. These observations demonstrate that quantized vortices actively reshape the normal-fluid dynamics, highlighting the necessity of fully coupled two-way simulations. These developments have motivated the formulation and numerical investigation of fully coupled two-fluid models that combine the VFM with the Navier--Stokes equations \cite{kivotides00,kivotides07,khomenko16,yui18,yui20,galantucci20,tang23,kobayashi24,yui25}.

A central issue in turbulence is the emergence of multiscale vortex structures and the associated energy cascade \cite{davidson15}. In classical turbulence, Kolmogorov's $-5/3$ power-law scaling is commonly interpreted in terms of the Richardson cascade. More recently, direct numerical simulations of the Navier--Stokes equations by Goto {\it et al.} have revealed that fully developed turbulence consists of hierarchical assemblies of anti-parallel vortex tubes spanning the inertial range \cite{goto17}. This finding provides a concrete physical picture of the cascade process in classical turbulence.

In quantum turbulence, however, whether such hierarchical structures can form remains an open question. Notably, even when quantized vortices possess the same rotational direction, they do not spontaneously aggregate to form a bundle. What mechanism, then, enables the formation of coherent quantized vortex bundles?

In this Letter, we perform fully two-way coupled simulations combining the vortex filament model with the Navier--Stokes equations to uncover the underlying mechanism. We demonstrate that the normal fluid drives the bundling of quantized vortices via mutual friction, establishing a direct connection between quantum turbulence and the hierarchical vortex-tube picture recently identified in classical turbulence.

{\it Coupled dynamics of two fluids.}---
We employ the VFM for quantized vortices and the Navier--Stokes equations for the normal fluid, with the two fluids coupled via mutual friction \cite{kivotides00,kivotides07}.
In the VFM, a quantized vortex is discretized into finite vortex filaments at discrete points ${\bm s}(\xi,t)$, representing the position vector with filament arc length $\xi$ at time $t$ \cite{schwarz85}.
The superfluid velocity ${\bm v}_{\text{s}}$ at a position ${\bm r}$ is obtained from the Biot--Savart law, integrating over all filaments along the path ${\cal L}$ \cite{adachi10} as ${\bm v}_\text{s} ({\bm r}) = (\kappa/4 \pi) \int _{\cal L} ({\bm s} - {\bm r}) \times \text{d} {\bm s} / |{\bm s} - {\bm r} |^{3} + {\bm v}_\text{s,b} + {\bm v}_\text{s,a}$, where $\kappa$ denotes the quantum of circulation, and ${\bm v}_\text{s,b}$ and ${\bm v}_\text{s,a}$ denote the velocities induced by the boundary conditions and the externally applied flow, respectively.
The Biot--Savart integration is accelerated by a factor of approximately 100 using the fast multipole method (FMM) \cite{yokota07,yokota09,yui25s} on a GPU.
The vortex filament velocity $\text{d} {\bm s}/\text{dt}$ obeys the equation of motion:
\begin{equation}
  \frac{\text{d} {\bm s}}{\text{dt}} = {\bm v}_\text{s} + \alpha {\bm s}' \times {\bm v}_\text{ns} - \alpha' {\bm s}' \times ({\bm s}' \times {\bm v}_\text{ns}),
  \label{vortex}
\end{equation}
where ${\bm v}_\text{ns}$ is the local relative velocity between the normal-fluid velocity ${\bm v}_\text{n}$ and superfluid velocity ${\bm v}_\text{s}$, $\alpha$ and $\alpha'$ are the temperature-dependent mutual friction coefficients, and ${\bm s}'$ is the unit tangent vector along the filament.
We use the 2W model \cite{schwarz78,schwarz85,barenghi83}, named in Ref.~\cite{tang23}, for the mutual friction coefficients to account for their temperature dependence. The model coefficients are obtained from experimental data for uniformly rotating superfluid helium \cite{hall56}. This model has been commonly used in one-way coupled simulations where normal-fluid distributions are prescribed and fixed temporally \cite{schwarz85,schwarz88,adachi10}, as well as in two-way coupled simulations \cite{yui18,yui20,yui25}. This choice is motivated by the fact that although the recently proposed self-consistent model \cite{galantucci20}---the S2W model named in Ref.~\cite{tang23}, which assumes a theoretical friction force along a vortex line without relying on empirical parameters---shows good agreement with experiments for the time evolution of quantized vortex ring propagation \cite{tang23}, it exhibits higher normal-fluid velocity fluctuations than observed experimentally \cite{kobayashi24}.
A detailed comparison between these models is provided in Sec. I of Supplemental Material 
\footnote{
See Sec. I of Supplemental Material for the results of the vortex line density, energy spectra, and vortex tubes obtained from the S2W model.
}.

The dynamics of the normal fluid is governed by the incompressible continuity equation $\nabla \cdot {\bm v}_\text{n} = 0$ and the incompressible Navier--Stokes equations with mutual friction ${\bm F}_\text{ns}$ \cite{donnelly91},
\begin{equation}
  \frac{ \partial {\bm v}_\text{n} }{\partial t} + ({\bm v}_\text{n} \cdot \nabla) {\bm v}_\text{n} = -\frac{1}{\rho_\text{n}} \nabla P + \nu_\text{n} \nabla^2 {\bm v}_\text{n} + \frac{{\bm F}_\text{ns} + {\bm F}_\text{ext}}{\rho_\text{n}},
  \label{eq:navier}
\end{equation}
where $P$, $\rho_\text{n}$, $\nu_\text{n}$, ${\bm F}_\text{ns}$, and ${\bm F}_\text{ext}$ denote the effective pressure, normal-fluid density, normal-fluid kinematic viscosity, mutual friction force, and external driving force, respectively.
The mutual friction force ${\bm F}_\text{ns} ({\bm x}) = (1/\Omega' ({\bm x})) \int_{ {\mathcal L}' ({\bm x})} {\bm f} (\xi)\text{d}\xi$ is calculated for each filament segment $\Delta \xi$ along the integral path ${\cal L}'$ within the normal-fluid mesh cell $\Omega'({\bm x})$ at position ${\bm x}$ \cite{kobayashi24}, with ${\bm f}(\xi)/\rho_\text{s} \kappa = \alpha {\bm s}' \times ( {\bm s}' \times  {\bm v}_\text{ns} ) + \alpha ' {\bm s}' \times {\bm v}_\text{ns}$ \cite{donnelly91}, where $\rho_\text{s}$ denotes the superfluid density.
A linear interpolation is used to evaluate the normal-fluid velocity ${\bm v}_\text{n}$ at the filament position ${\bm s}$ in Eq. (\ref{vortex}).

{\it Numerical conditions.}---
A cubic computational domain of size $D_x = D_y = D_z = 1.0~\mathrm{mm}$ is employed with periodic boundary conditions in all directions.
The spatial resolution along the discrete vortex filaments is set within the range $4.0 \times 10^{-4}~\mathrm{cm} < \Delta \xi < 1.2 \times 10^{-3}~\mathrm{cm}$, with a minimum resolution of $\Delta \xi_\text{min} = 4.0 \times 10^{-4}~\mathrm{cm}$.
For the normal fluid, a uniform grid of $N_x = N_y = N_z = 128$ points is used.
The time step for both fluid components is set to $\Delta t = 2.0 \times 10^{-5}~\mathrm{s}$.
Vortex reconnections are implemented algorithmically whenever two vortex lines approach closer than $\Delta \xi_\text{min}$, and vortex loops shorter than $5 \Delta \xi_\text{min}$ are removed.

We apply a Taylor--Green-type external driving force \cite{goto17},
${\bm F}_\text{ext}/\rho_\text{n} = \left( -F \sin 2 \pi x/D_x \cos 2 \pi y/D_y, \right.$ $\left. F \cos 2 \pi x/D_x \sin 2 \pi y/D_y, 0 \right)$, 
where the force amplitude is set to $F = 0.025~\mathrm{m/s^2}$.

Equation (\ref{eq:navier}) for the normal fluid is discretized using a second-order accurate finite difference method. 
The simplified Marker-and-Cell (SMAC) method is used to couple the velocity and pressure, and the fast Fourier transform is applied to solve the pressure Poisson equation.
Time integration is performed using the second-order Adams--Bashforth scheme for the normal fluid and the fourth-order Runge--Kutta scheme for the superfluid.
The temperature is primarily set to $T = 1.9~\mathrm{K}$, with physical parameters adopted from Ref.~\cite{barenghi83}.
Simulations are initialized with eight randomly oriented quantized vortex loops in a normal fluid at rest.


\begin{figure}
  \centering
  \includegraphics[width=1.0\linewidth]{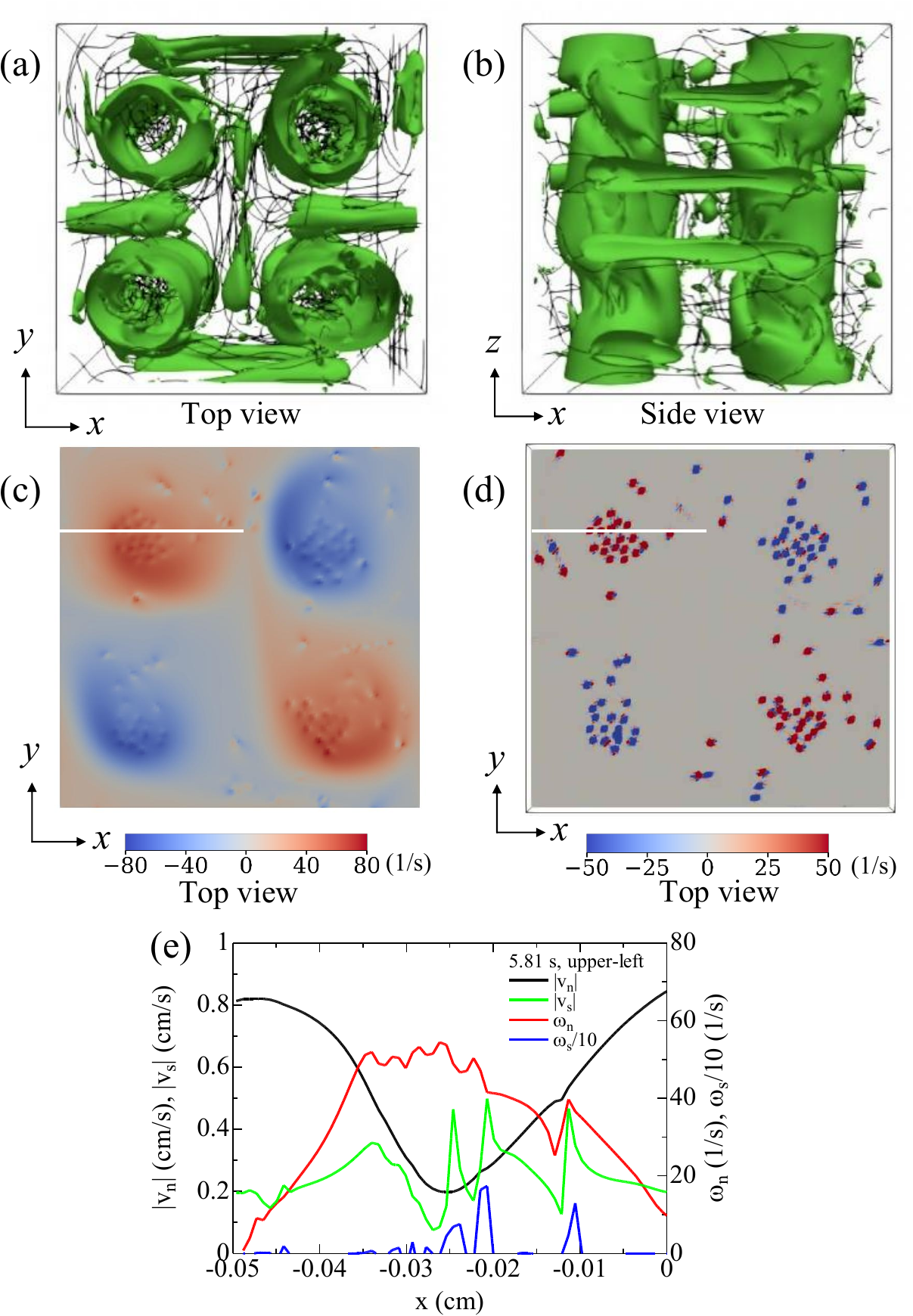}
  \caption
  {
  Instantaneous distributions of quantized vortices (black lines) and normal-fluid vortex tubes (green isosurfaces for the second invariant of the velocity gradient tensor $Q = 1000~\mathrm{s^{-2}}$) at $t = 5.81~\mathrm{s}$ in (a) top view and (b) side view. 
Panels (c) and (d) show the corresponding top views of the $z$-component of normal-fluid vorticity $\omega_{\text{n},z}$ and superfluid vorticity $\omega_{\text{s},z}$, respectively, with white solid lines indicating the extraction line for spatial profiles. 
(e) One-dimensional spatial profiles of normal-fluid velocity magnitude $|{\bm v}_\text{n}|$ (black), superfluid velocity magnitude $|{\bm v}_\text{s}|$ (green), normal-fluid vorticity $\omega_{\text{n},z}$ (red), and scaled superfluid vorticity $\omega_{\text{s},z}/10$ (blue) along the white line in (c) and (d).
  }
  \label{fig:qomega-vs}
\end{figure}

{\it Results and discussion.}---
Figures \ref{fig:qomega-vs}(a) and \ref{fig:qomega-vs}(b) display instantaneous distributions of quantized vortices and normal-fluid vortex tubes visualized by the second invariant of the velocity gradient tensor, $Q = 1000~\mathrm{s^{-2}}$.
The region with $Q > 0$ indicates where vorticity dominates over strain, defined as $Q = (W_{ij}W_{ij} - S_{ij}S_{ij})/2$ in terms of the vorticity tensor $W_{ij} = (\partial u_i/\partial x_j - \partial u_j/\partial x_i)/2$ and the velocity strain-rate tensor $S_{ij} = (\partial u_i/\partial x_j + \partial u_j/\partial x_i)/2$.
The dynamics can be seen in the movies in Supplemental Material 
\footnote
{
Movies are provided in Supplemental Material.
}.
Driven by the continuously applied Taylor--Green-type external force, four large normal-fluid vortex tubes (parent vortices) are generated, which subsequently produce child vortices perpendicular to the parent vortices \cite{goto17}, as clearly seen in the side view in Fig.~\ref{fig:qomega-vs}(b).
Quantized vortices are stretched by the normal fluid via mutual friction, eventually transitioning to quantum turbulence.
The quantized vortices are rolled up and form bundles inside the normal-fluid vortex tubes.

Figures \ref{fig:qomega-vs}(c) and \ref{fig:qomega-vs}(d) present cross-sectional top views of the normal-fluid and superfluid $z$-vorticity distributions (hereafter, vorticity simply denotes the $z$-component).
Red and blue indicate positive and negative vorticity, corresponding to counter-clockwise and clockwise rotations, respectively.
The quantized vortices align in the same orientation as the normal fluid vorticity. 
Many quantized vortices are confined within a parent normal-fluid vortex tube whereas at most a single quantized vortex is observed within a child vortex tube.
This means that the number of bundled quantized vortices depends on the normal-fluid circulation, scaling with the core-region size $r$ as $r^{4/3}$ (see Sec.II of Supplemental Material 
\footnote{
See Sec.~II of Supplemental Material for the derivation of the dependence of the number of bundled quantized vortices on the size $r$.
}).

Figure \ref{fig:qomega-vs}(e) shows the profiles of normal-fluid velocity $|{\bm v}_\text{n}|$ (black) and superfluid velocity $|{\bm v}_\text{s}|$ (green), as well as normal-fluid vorticity $\omega_{\text{n},z}$ (red) and superfluid vorticity $\omega_{\text{s},z}/10$ (blue), evaluated along the white line in Figs.~\ref{fig:qomega-vs}(c) and \ref{fig:qomega-vs}(d) at $t = 5.81~\mathrm{s}$.
The normal-fluid velocity profile exhibits solid-body rotation in the core region ($-0.035~\mathrm{cm} < x < -0.015~\mathrm{cm}$), resulting in a nearly uniform normal-fluid vorticity.
The superfluid velocity in the core region stays close to the normal-fluid velocity owing to mutual friction.
The superfluid vorticity exists discretely, reflecting the distribution of individual quantized vortices as seen in Fig.~\ref{fig:qomega-vs}(d).
Such bundled vortex configurations have been theoretically studied in superfluid under solid-body rotation \cite{campbell79,tsubota02} and observed experimentally \cite{yarmchuk82,madison01}.

\begin{figure}
  \centering
  \includegraphics[width=1.0\linewidth]{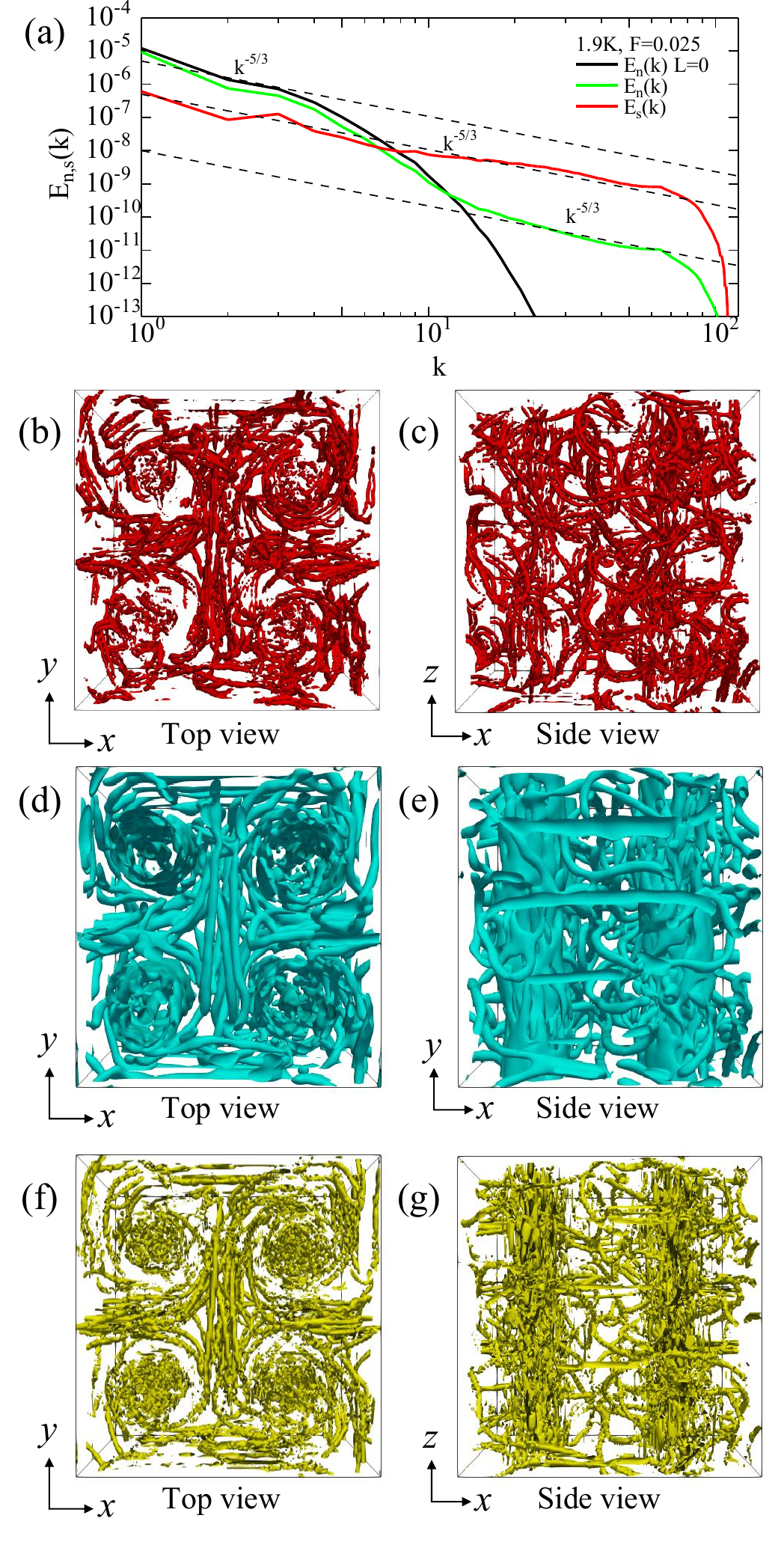}
  \caption
  {
(a) Time-averaged ($5 \le t \le 20~\mathrm{s}$) energy spectra of the uncoupled normal fluid calculated from the Navier--Stokes equation, Eq.~\eqref{eq:navier}, alone (black) and the two-way coupled system for the superfluid $E_\text{s}(k)$ (red) and normal fluid $E_\text{n}(k)$ (green). The dashed lines indicate a $k^{-5/3}$ power-law slope. (b)--(g) Instantaneous distributions of vortex tubes filtered by wavenumber ranges at $t = 5.81~\mathrm{s}$: (b) top view and (c) side view of high-wavenumber ($k > 15$) normal-fluid vortex tubes (red: $Q = 50~\mathrm{s^{-2}}$); (d) top view and (e) side view of low-wavenumber ($k \le 15$) superfluid vortex tubes (light blue: $Q = 500~\mathrm{s^{-2}}$); and (f) top view and (g) side view of high-wavenumber ($k > 15$) superfluid vortex tubes (yellow: $Q = 500~\mathrm{s^{-2}}$).
  }
  \label{fig:enespe-qomga-h}
\end{figure}

Figure \ref{fig:enespe-qomga-h}(a) presents the time-averaged ($5 \le t \le 20\text{ s}$) energy spectra of the uncoupled normal fluid calculated from the Navier--Stokes equation, Eq.~\eqref{eq:navier}, alone (black) and the two-way coupled system for the superfluid (red) and normal fluid (green).
The energy spectrum without quantized vortices (black) decays at high wavenumbers due to the viscous dissipation. 
At low wavenumbers, the range exhibiting a $k^{-5/3}$ power law is narrow owing to the modest forcing magnitude.
In contrast, the energy spectrum of the superfluid component (red) closely follows a $k^{-5/3}$ power law across a wider range. 
For the normal-fluid component (green), the energy spectrum at low wavenumbers remains almost unchanged from the uncoupled case without quantized vortices. 
However, beyond the wavenumber corresponding to the mean inter-vortex spacing, the normal-fluid energy is enhanced by drag forces exerted by quantized vortices via mutual friction, subsequently recovering a $k^{-5/3}$ power-law behavior.

Figures \ref{fig:enespe-qomga-h}(b) and \ref{fig:enespe-qomga-h}(c) exhibit the instantaneous normal-fluid vortex tubes of high wavenumbers ($k > 15$).
The spatial distribution of high-wavenumber vortex tubes closely matches that of the quantized vortices.
These high-wavenumber vortex tubes are induced by the quantized vortices via mutual friction and are located adjacent to the vortex lines, consistent with observations in previous studies \cite{yui20,galantucci20,tang23,kobayashi24}.
In contrast, the distributions of low-wavenumber normal-fluid vortex tubes coincide with those in Figs.~\ref{fig:qomega-vs}(a) and \ref{fig:qomega-vs}(b).
The smaller threshold $Q = 50~\mathrm{s^{-2}}$ used for the high-wavenumber tubes reflects weaker vorticity compared to $Q = 1000~\mathrm{s^{-2}}$ for the low-wavenumber vortex tubes.

On the other hand, as shown in Figs.~\ref{fig:enespe-qomga-h}(d)--\ref{fig:enespe-qomga-h}(g), the instantaneous distributions of low-wavenumber superfluid vortex tubes are remarkably similar to those of high-wavenumber ones, with $Q = 500~\mathrm{s^{-2}}$ applied to both wavenumber ranges.
The superfluid vortex tubes are consistently observed around the quantized vortices, with the primary difference lying in their thickness: high and low wavenumbers correspond to thin and thick vortex tubes, respectively.
This hierarchical vortex-tube configuration in the superfluid yields an energy spectrum with the same power-law exponent extending continuously from low to high wavenumbers, as shown in Fig.~\ref{fig:enespe-qomga-h}(a).
Consequently, such superfluid high-wavenumber fluctuations at scales smaller than the mean inter-vortex spacing excite high-wavenumber normal-fluid motions via mutual friction, resulting in another $k^{-5/3}$ spectrum.

\begin{figure}
  \centering
  \includegraphics[width=0.7\linewidth]{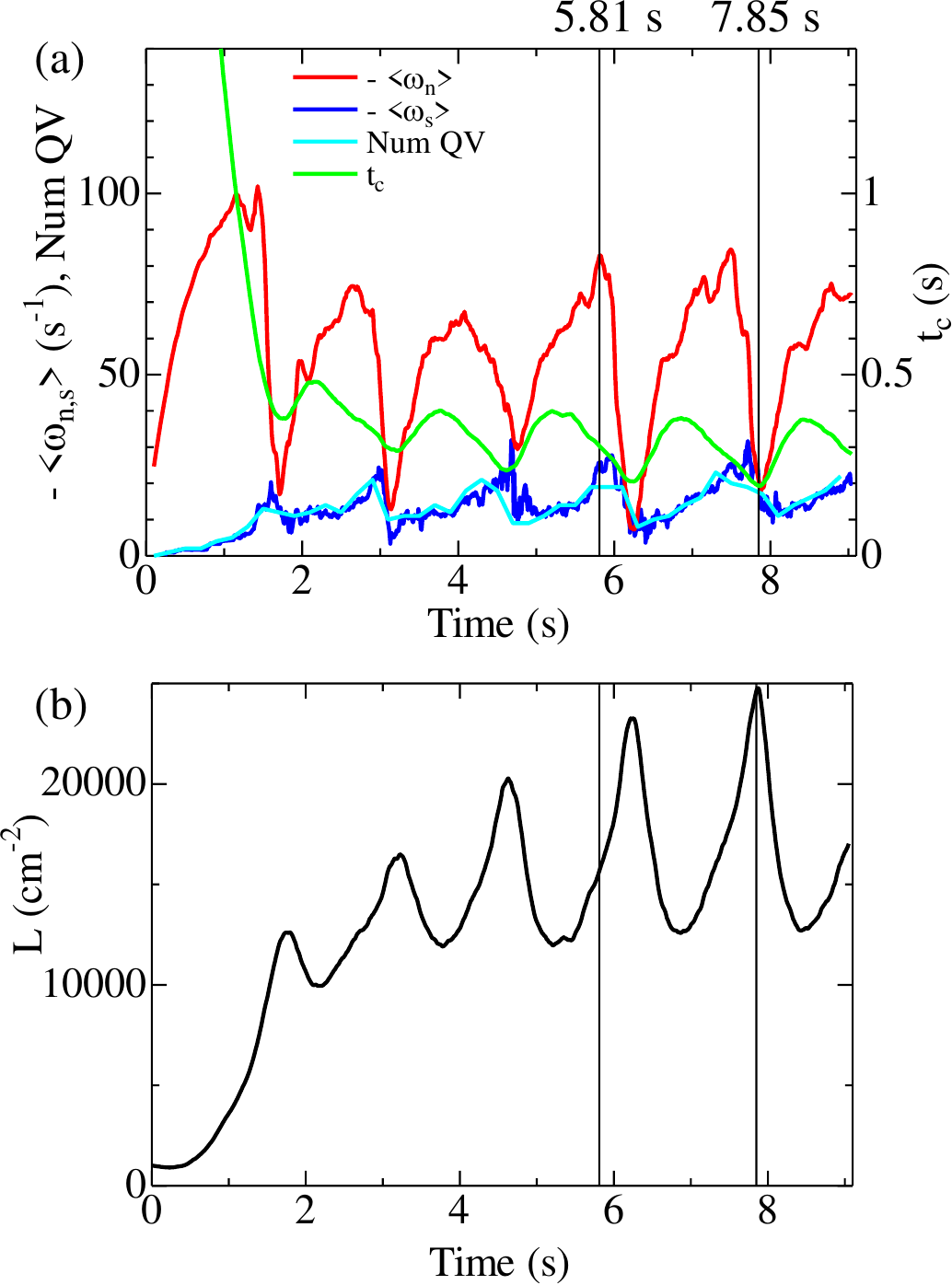}
  \caption
  {
  (a) Time evolutions of the spatial-averaged normal-fluid vorticity $\langle \omega_\text{n} \rangle$ (red), superfluid vorticity $\langle \omega_\text{s} \rangle$ (blue), number of quantized vortices (light blue) evaluated for $Q > 1000~\mathrm{s^{-2}}$, and coupling timescale $t_\text{c}$ (green), and (b) time evolution of the total vortex line density $L$.
Vertical lines indicate $t = 5.81~\mathrm{s}$ and $7.85~\mathrm{s}$.
  }
  \label{fig:t-ome-L}
\end{figure}

Next, we focus on the flow dynamics within the bottom-left quadrant of Figs. \ref{fig:qomega-vs}(c) and \ref{fig:qomega-vs}(d).
Figure~\ref{fig:t-ome-L}(a) depicts the time evolutions of the domain-averaged normal-fluid and superfluid vorticities, $-\langle \omega \rangle_{\text{n,s}}$, within the normal-fluid region of $Q > 1000\text{ s}^{-2}$ in this quadrant, where $\langle \cdot \rangle$ denotes the spatial average. 
Here, $\langle \omega \rangle_{\text{n,s}} = \Gamma_{\text{n,s}} / \int_{S'} dx\,dy$, where $\Gamma_{\text{n,s}} = \int_{S'} \omega_{\text{n,s}}(\bm{x})dx\,dy$ represents the circulation and $S'$ denotes the area satisfying $Q > 1000\text{ s}^{-2}$.
The magnitude of the normal-fluid vorticity $-\langle \omega \rangle_{\text{n}}$ is consistently larger than that of the superfluid vorticity $-\langle \omega \rangle_{\text{s}}$.
Note that the temporal evolution of $-\Gamma_{\text{n,s}}/\kappa$ closely tracks that of $-\langle \omega \rangle_{\text{n,s}}$ (not shown here).
Notably, $-\Gamma_{\text{s}}/\kappa$ agrees well with the number of quantized vortices (Num QV: light blue) counted inside $Q > 1000\text{ s}^{-2}$.
The number of quantized vortices (Num QV) increases with $-\langle \omega \rangle_{\text{n}}$, exhibiting a distinct time lag at its peak.
As shown in Fig.~\ref{fig:t-ome-L}(b), the total vortex line density $L$ reaches its peak when the vorticities of both fluids drop to their minimum.
The instantaneous vortex distributions at $t = 7.85\text{ s}$ are provided in Sec. III of Supplemental Material
\footnote{
See Sec. III of Supplemental Material for the instantaneous vortex distributions at $t = 7.85\text{ s}$.
}.
The coupling time scale $t_{\text{c}} = 1/(\alpha \kappa L)$ \cite{vinen00} (green in Fig.~\ref{fig:t-ome-L}(a)), derived in Sec. IV of Supplemental Material 
\footnote{
See Sec. IV of Supplemental Material for the derivation of the coupling time scale.
}, reaches its minimum at this vorticity minimum, providing strong mutual friction coupling.

\begin{figure}
  \centering
  \includegraphics[width=1.0\linewidth]{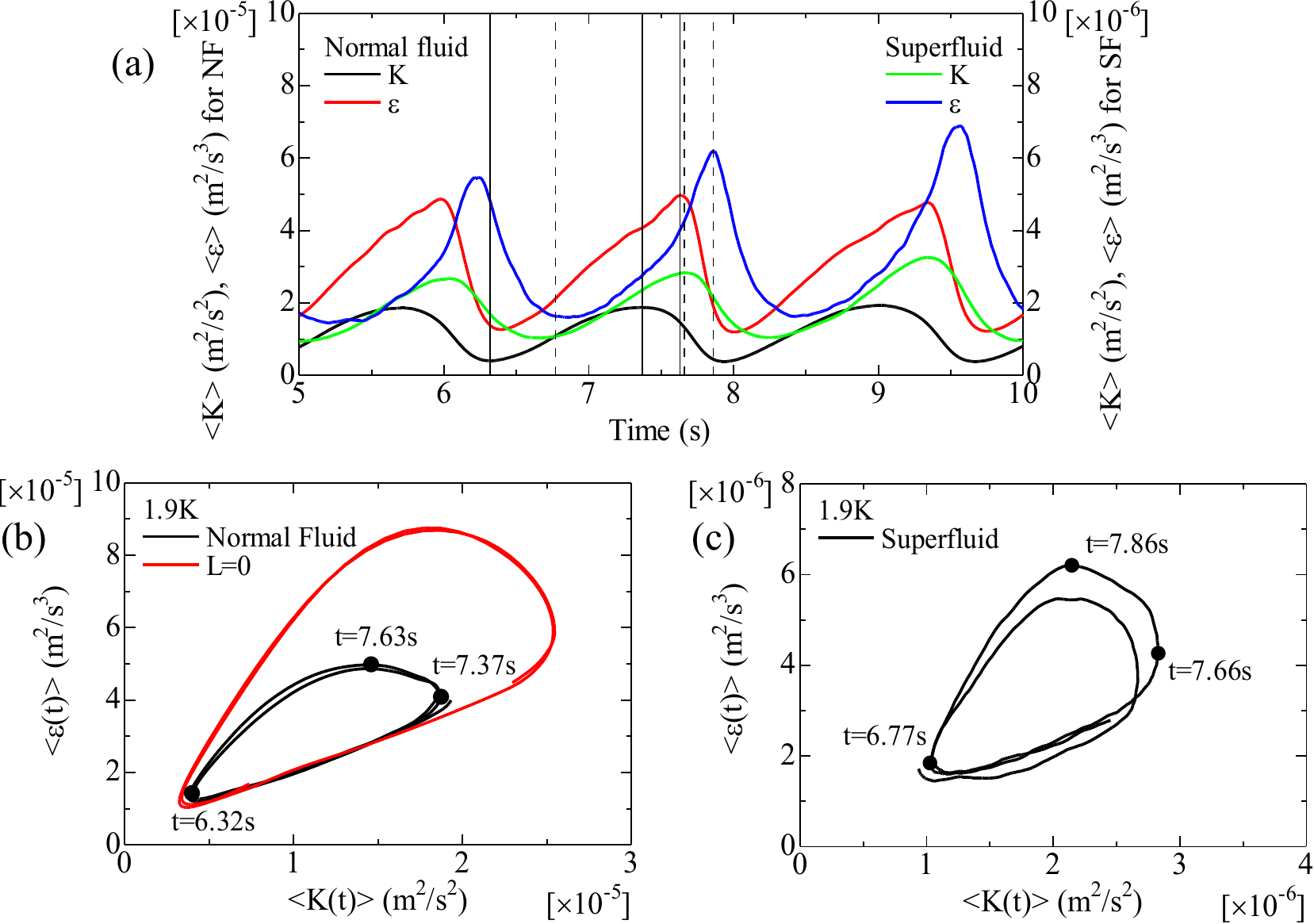}
  \caption
  {
  (a) Time evolutions of kinetic energy $\langle K \rangle$ (normal fluid: black, superfluid: green) and energy dissipation rate $\langle \varepsilon \rangle$ (normal fluid: red, superfluid: blue). Vertical solid and dashed lines indicate corresponding times for the normal fluid ($t = 6.32$, $7.37$, and $7.63~\mathrm{s}$) and superfluid ($t = 6.77$, $7.66$, and $7.86~\mathrm{s}$), respectively.
  (b) Phase-space trajectories of $\langle K(t) \rangle$ versus $\langle \varepsilon(t) \rangle$ for the normal fluid without (red) and with (black) quantized vortices.
  (c) Phase-space trajectory of $\langle K(t) \rangle$ versus $\langle \varepsilon(t) \rangle$ for the superfluid.
  }
  \label{fig:limitcycle}
\end{figure}

As shown in Fig.~\ref{fig:limitcycle}(a), both fluids simultaneously reach their minimal values of kinetic energy $\langle K \rangle$ and energy dissipation rate $\langle \varepsilon \rangle$.
This synchronicity also holds for their maxima, where kinetic energy peaks first, followed by the energy dissipation rate. 
Furthermore, a temporal delay exists between the two fluids.
The limit cycle in the kinetic energy--energy dissipation rate plane for the normal fluid with quantized vortices (black) shrinks relative to the uncoupled case (red) [Fig. \ref{fig:limitcycle}(b)].
The reduction in normal-fluid energy associated with the shrinking of the limit cycle is transferred to the superfluid via mutual friction, as shown in Fig. \ref{fig:limitcycle}(c), where the superfluid energy dissipation rate $\varepsilon_\text{s}$ is calculated by $\varepsilon_\text{s} = \nu' \kappa^2 L^2$ assuming $\nu'/\kappa = 0.1$.

As demonstrated in classical (Navier--Stokes) turbulence \cite{goto17}, when turbulence is sustained by a steady Taylor--Green-type forcing, a hierarchy of anti-parallel normal-fluid vortices evolves quasiperiodically with temporal fluctuations. A finite time lag in the active transfer from large-scale vortices to small-scale vortices leads to a scale-local energy cascade from the forcing scale to the dissipative scale over a finite time. In the present two-way coupled system, the time lag in kinetic energy between the normal fluid and the superfluid is governed by the strength of mutual friction, whereas the time lag in the superfluid between its kinetic energy and energy dissipation rate drives the superfluid energy cascade, yielding a $k^{-5/3}$ energy spectrum in the superfluid.

\begin{figure}
  \centering
  \includegraphics[width=1.0\linewidth]{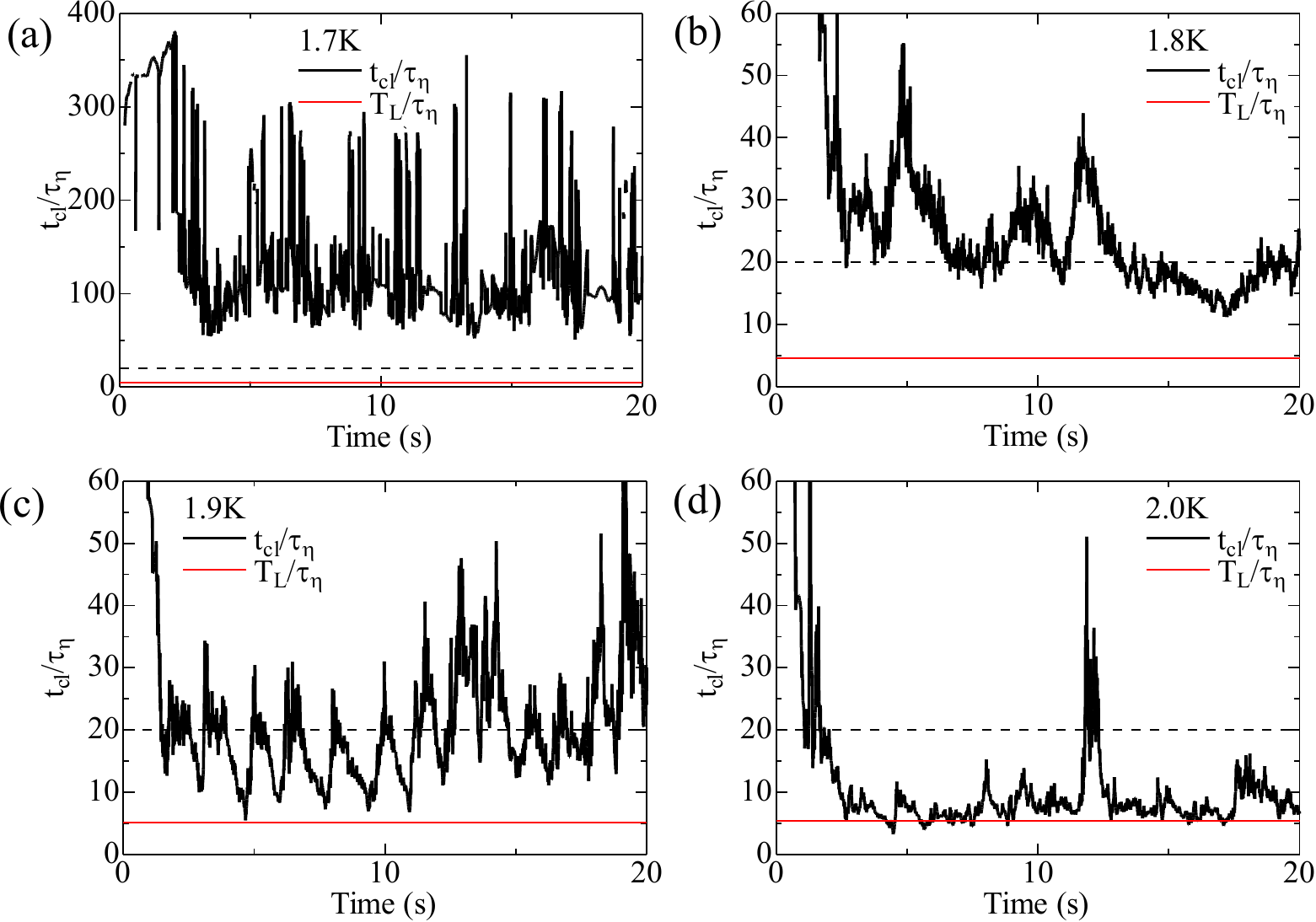}
  \caption
  {
  Time evolutions of the ratio of the local coupling time scale $t_\mathrm{cl} = 1/(\alpha \kappa L_{Q>1000})$ to the Kolmogorov dissipation time scale $\tau_\eta$ at (a) $T = 1.7~\mathrm{K}$, (b) $1.8~\mathrm{K}$, (c) $1.9~\mathrm{K}$, and (d) $2.0~\mathrm{K}$ [black dashed line: $20$; red line: $T_L / \tau_\eta$, where $T_L$ is the eddy turnover time].
  }
  \label{fig:timescale}
\end{figure}

The time scale associated with the coupling between the superfluid and normal fluid has been a major open question since Vinen's landmark study \cite{vinen00}.
The underlying physics of this coupling can be revealed only through two-way coupled two-fluid simulations, which we systematically investigate as a function of temperature.
Figure \ref{fig:timescale} shows the time evolutions of the ratio of the local coupling time scale $t_\mathrm{cl} = 1/(\alpha \kappa L_{Q>1000})$ to the Kolmogorov dissipation time scale $\tau_\eta$, where $L_{Q>1000}$ represents the local vortex line density evaluated for $Q > 1000\text{ s}^{-2}$ within the bottom-left quadrant.
Vinen \cite{vinen00} proposed that the condition for the two fluids to lock together as a single classical fluid is $t_\mathrm{c} \ll T_L$, where $T_L$ is the eddy turnover time.
Comparing $t_\mathrm{cl}/\tau_\eta$ with $T_L/\tau_\eta$ shows that most of the simulation time corresponds to the conventional decoupled regime ($t_\mathrm{cl} > T_L$).
Nevertheless, as demonstrated above, bundled quantized vortices clearly form within the normal-fluid vortex tubes.
We find that the threshold for this dynamic bundle formation is $t_\mathrm{cl}/\tau_\eta \approx 20$; i.e., the condition $t_\mathrm{cl}/\tau_\eta < 20$ governs the local bundling inside the normal-fluid vortex tubes.
As the temperature increases, the duration satisfying $t_\mathrm{cl}/\tau_\eta < 20$ significantly expands, demonstrating that the two-fluid coupling is locally enhanced at higher temperatures due to the increase in the mutual friction coefficient $\alpha$.

{\it Conclusions.}---%
We have demonstrated the two-way coupled dynamics of quantized vortex bundle formation in spatially periodic normal-fluid turbulence.
Even when quantized vortices rotate in the same direction, they do not spontaneously aggregate to form a bundle. The present study reveals that the normal fluid drives the bundling of quantized vortices via mutual friction.
Quantized vortices are rolled up via mutual friction and form bundles inside normal-fluid vortex tubes.
The number of bundled quantized vortices is determined by the core region where the normal-fluid vorticity is uniform and scales directly with the normal-fluid circulation, corresponding to the tube thickness.
This demonstrates that the spatial hierarchy observed in classical turbulence is inherited by quantized vortex bundles.

Crucially, the superfluid exhibits a $k^{-5/3}$ energy spectrum and actively excites high-wavenumber normal-fluid fluctuations via mutual friction, manifesting as normal-fluid vortex tubes generated adjacent to quantized vortices. In contrast, the spatial structure of the superfluid forms a distinct nested configuration around the quantized vortices, where low-wavenumber components form thick outer vortex tubes, while high-wavenumber components represent fine inner vortex tubes embedded inside them.

Furthermore, the bundle formation process recurs periodically.
The temporal evolutions of kinetic energy and energy dissipation rate in the superfluid exhibit a distinct time lag relative to those in the normal fluid.
This time lag drives the superfluid energy cascade, yielding the observed $k^{-5/3}$ spectrum.
The duration for which a bundle remains trapped inside a normal-fluid vortex tube is governed by the regime satisfying the local coupling condition $t_\mathrm{cl}/\tau_\eta < 20$.
Consequently, as the regime satisfying this time-scale condition expands with increasing temperature, the vortex bundles exhibit enhanced temporal stability at higher temperatures.

Finally, these findings offer valuable insights into two-fluid turbulence dynamics in experiments and provide key implications for the power-law decay of vortex line density in decaying turbulence. Moreover, the observed excitation of high-wavenumber normal-fluid motions by the superfluid strongly suggests that quantum turbulence can accelerate the transition to turbulence in the normal fluid.

\begin{acknowledgments}
H. K. was supported by JSPS KAKENHI Grant No. JP23K22674. 
M. T. was supported by JSPS KAKENHI Grant No. JP22H05139 and No. JP22H05131. 
This work used computational resources of TSUBAME4.0 provided by Institute of Science Tokyo through the HPCI System Research Project (Project ID: hp250129, hphp260233).
\end{acknowledgments}

\section*{Supplemental Material}

\renewcommand{\theequation}{S.\arabic{equation}}
\renewcommand{\thefigure}{S.\arabic{figure}}
\renewcommand{\thetable}{S.\arabic{table}}
\renewcommand{\thesubsection}{\Roman{subsection}}
\setcounter{equation}{0}
\setcounter{figure}{0}
\setcounter{table}{0}
\setcounter{subsection}{0}

\subsection{Vortex line density, energy spectra, and vortex tubes obtained from the S2W model}

\begin{figure}[h]
  \centering
  \includegraphics[width=0.9\linewidth]{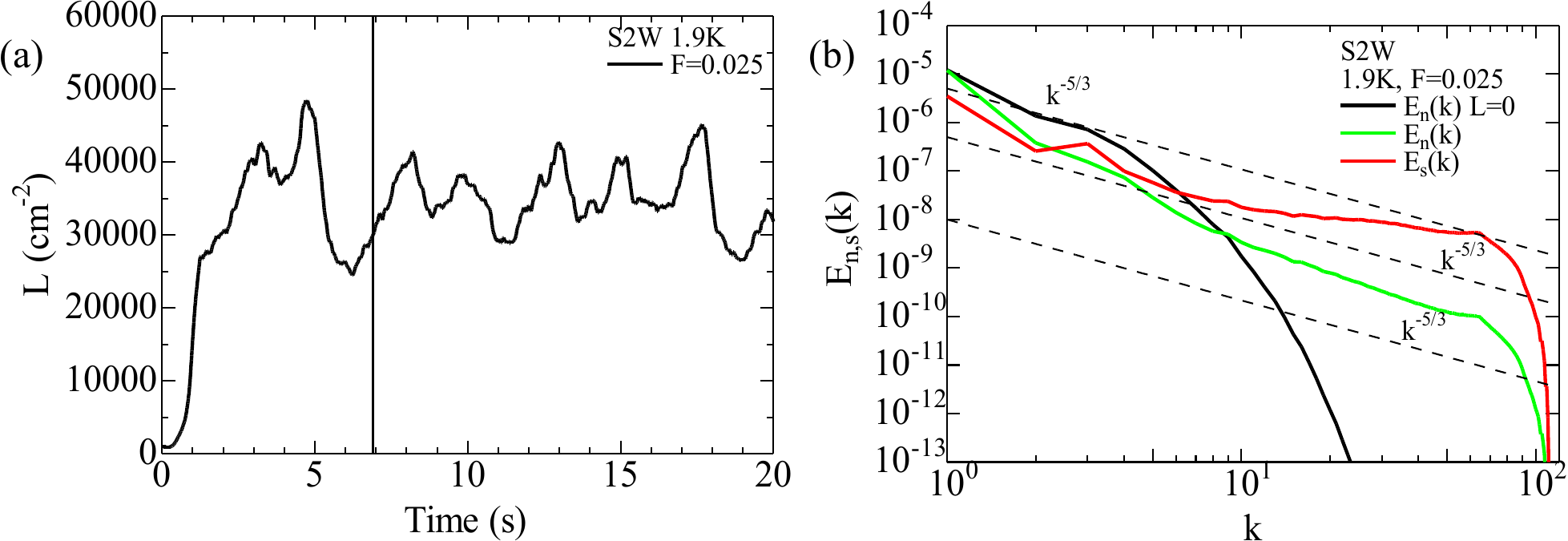}
  \caption
  {
  (a) Time evolution of the vortex line density (the vertical dashed line marks $t = 6.90\text{ s}$) and (b) time-averaged ($5 \le t \le 20\text{ s}$) energy spectra of the uncoupled normal fluid calculated from the Navier--Stokes equations alone (black line), and the two-way coupled system for the superfluid (red line) and normal fluid (green line) at 1.9 K and $F = 0.025\text{ m/s}^2$ using the S2W model. Dashed lines indicate a $k^{-5/3}$ slope
  }
  \label{fig:t-L-enespe-G19K-25F}
\end{figure}

We conducted two-way coupled simulations using the S2W model \cite{galantucci20} at 1.9 K and $F = 0.025\text{ m/s}^2$ to compare with results obtained using the 2W model \cite{schwarz78,schwarz85,barenghi83}.
The time evolution of the vortex line density, shown in Fig. \ref{fig:t-L-enespe-G19K-25F}(a), exhibits a periodic yet somewhat chaotic behavior.
The vortex line density calculated with the S2W model is significantly higher than that with the 2W model.
As shown in Fig. \ref{fig:t-L-enespe-G19K-25F}(b), the time-averaged ($5 \le t \le 20\text{ s}$) energy spectrum of the two-way coupled system for the superfluid (red line) deviates from Kolmogorov's $k^{-5/3}$ power-law slope at high wavenumbers, indicating the presence of strong velocity fluctuations in this high-wavenumber regime.
These superfluid fluctuations strongly excite high-wavenumber normal-fluid fluctuations, resulting in an elevated normal-fluid energy spectrum (green line).

\begin{figure}[h]
  \centering
  \includegraphics[width=1.0\linewidth]{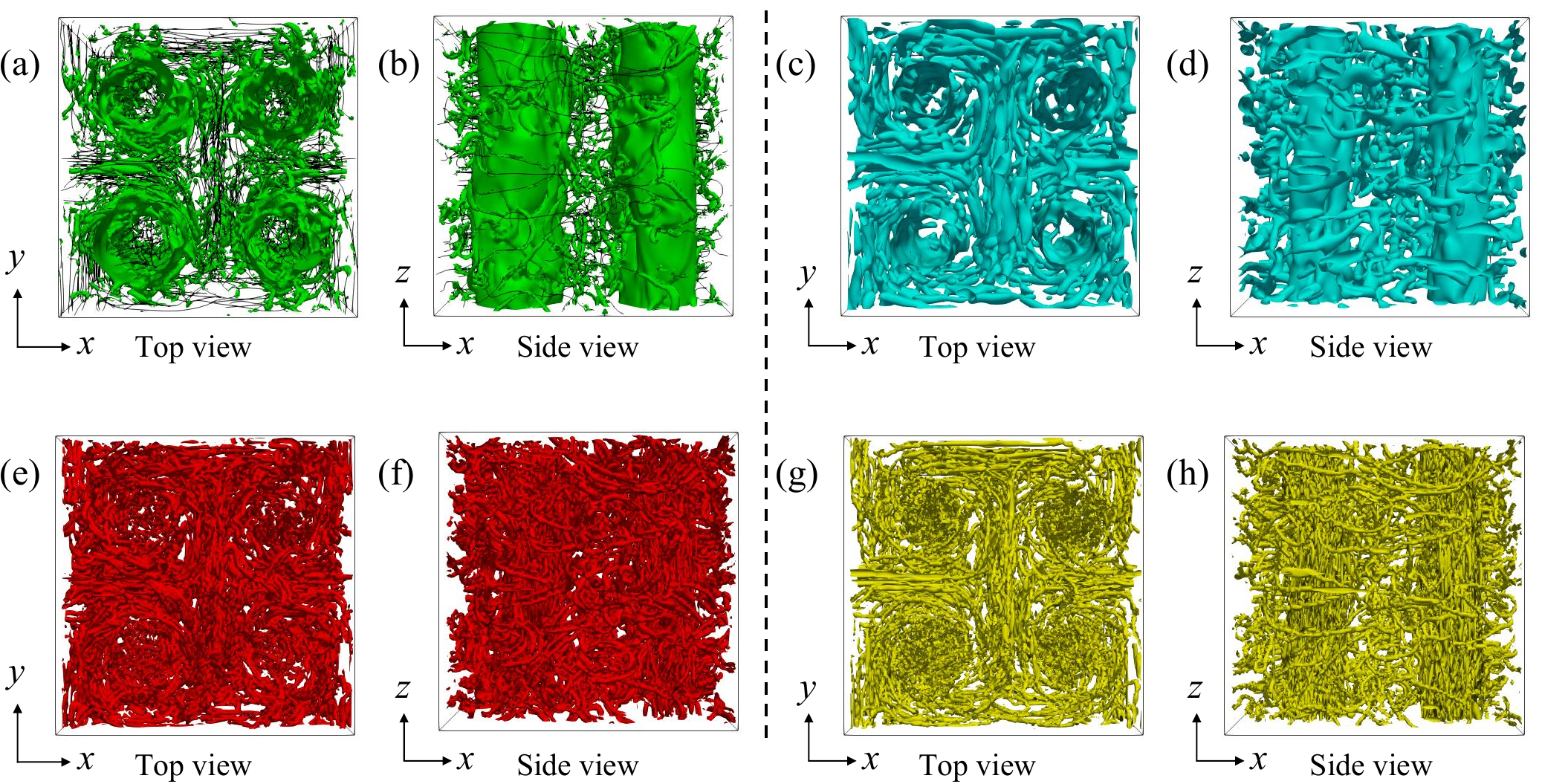}
  \caption
  {
  Instantaneous distributions of quantized vortices (black lines) and vortex tubes at $t = 6.90\text{ s}$ for 1.9 K and $F = 0.025\text{ m/s}^2$ using the S2W model: (a) top view and (b) side view of low-wavenumber ($k \le 15$) normal-fluid vortex tubes (green: $Q = 1000\text{ s}^{-2}$); (c) top view and (d) side view of low-wavenumber ($k \le 15$) superfluid vortex tubes (light blue: $Q = 500\text{ s}^{-2}$); (e) top view and (f) side view of high-wavenumber ($k > 15$) normal-fluid vortex tubes (red: $Q = 50\text{ s}^{-2}$); and (g) top view and (h) side view of high-wavenumber ($k > 15$) superfluid vortex tubes (yellow: $Q = 500\text{ s}^{-2}$).
  }
  \label{fig:qns-G19K-25F-690}
\end{figure}

Furthermore, as illustrated in Fig. \ref{fig:qns-G19K-25F-690}, both the high-wavenumber normal-fluid vortex tubes [(e) and (f)] and superfluid vortex tubes [(g) and (h)] are more concentrated than those obtained with the 2W model.


\subsection{Number of bundled quantized vortices}

The velocity circulation $\Gamma$ along a closed loop for a velocity field $\bm{u}$ is defined as
\begin{equation}
\Gamma = \oint \bm{u} \cdot d\bm{l} = \int_S \bm{\omega} \cdot d\bm{S} = 2 \Omega S = n S \kappa,
\end{equation}
where $d\bm{l}$ denotes the differential line element, $S$ is the cross-sectional area, $\bm{\omega} = \nabla \times \bm{u}$ is the vorticity, $\Omega$ is the mean vorticity, $n$ is the areal number density of vortices, and $\kappa$ is the quantum of circulation.
In a superfluid, the Feynman relation gives $n = 2\Omega / \kappa$.

In normal-fluid turbulence, the energy dissipation rate $\varepsilon$ at scale $r$ is approximated as $\varepsilon \sim u^3 / r$, giving $u \sim \varepsilon^{1/3} r^{1/3}$.
Consequently, the vorticity scales as $\omega \sim du/dr \sim \varepsilon^{1/3} r^{-2/3}$, and the circulation scales as $\Gamma \sim u r \sim \varepsilon^{1/3} r^{4/3}$.
The number of bundled quantized vortices within a region of scale $r$ is estimated as
\begin{equation}
N_{\text{QV}} = n S = \frac{\Gamma}{\kappa} \sim \frac{\varepsilon^{1/3}}{\kappa} r^{4/3}.
\label{eq:nqv_scaling}
\end{equation}
Equation~\eqref{eq:nqv_scaling} indicates that the number of bundled quantized vortices depends on scales as $r^{4/3}$, which is a defining feature of vortex bundles formed within normal-fluid vortex tubes.

\subsection{Normal fluid and superfluid vortex tubes at 7.85 s}

\begin{figure}[h]
  \centering
  \includegraphics[width=1.0\linewidth]{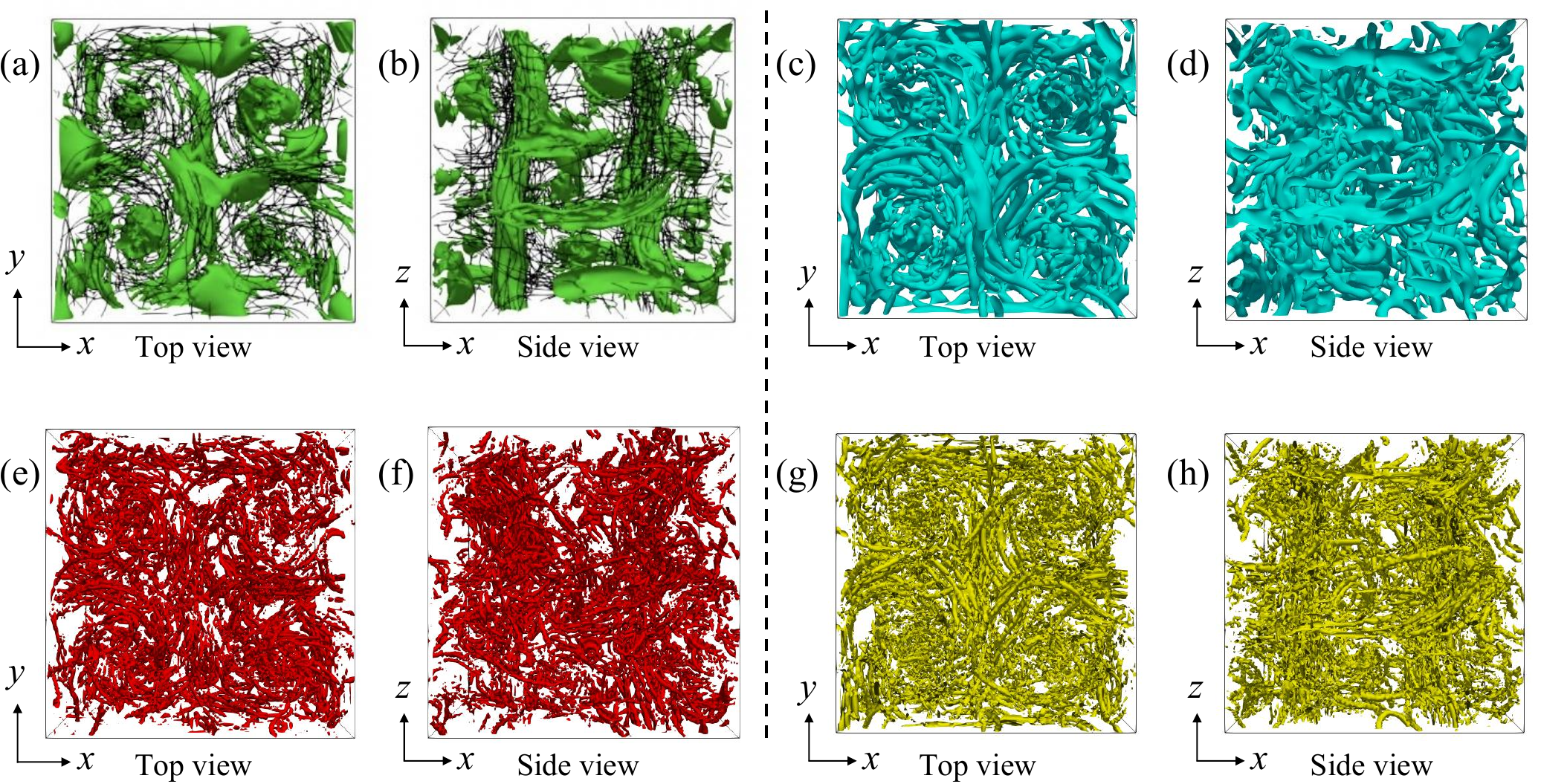}
  \caption
  {
  Instantaneous distributions of quantized vortices (black lines) and vortex tubes at $t = 7.85\text{ s}$ for 1.9 K and $F = 0.025\text{ m/s}^2$ using the 2W model: (a) top view and (b) side view of low-wavenumber ($k \le 15$) normal-fluid vortex tubes (green: $Q = 1000\text{ s}^{-2}$); (c) top view and (d) side view of low-wavenumber ($k \le 15$) superfluid vortex tubes (light blue: $Q = 500\text{ s}^{-2}$); (e) top view and (f) side view of high-wavenumber ($k > 15$) normal-fluid vortex tubes (red: $Q = 50\text{ s}^{-2}$); and (g) top view and (h) side view of high-wavenumber ($k > 15$) superfluid vortex tubes (yellow: $Q = 500\text{ s}^{-2}$).
  }
  \label{fig:qns-785}
\end{figure}

At $t = 7.85\text{ s}$, which corresponds to a peak in the vortex line density, Figs. \ref{fig:qns-785}(a) and \ref{fig:qns-785}(b) show the instantaneous distributions of quantized vortices (black lines) and low-wavenumber ($k \le 15$) normal-fluid vortex tubes (green: $Q = 1000\text{ s}^{-2}$).
The corresponding high-wavenumber ($k > 15$) normal-fluid vortex tubes (red: $Q = 50\text{ s}^{-2}$) are shown in Figs. \ref{fig:qns-785}(e) and \ref{fig:qns-785}(f), illustrating that high-wavenumber vortex tubes are generated adjacent to the quantized vortices.

For the superfluid component, Figs. \ref{fig:qns-785}(c) and \ref{fig:qns-785}(d) display the low-wavenumber ($k \le 15$) vortex tubes (light blue: $Q = 500\text{ s}^{-2}$), while Figs. \ref{fig:qns-785}(g) and \ref{fig:qns-785}(h) visualize the high-wavenumber ($k > 15$) vortex tubes (yellow: $Q = 500\text{ s}^{-2}$).
These superfluid vortex tubes are formed around the quantized vortices.

\subsection{Derivation of the two-fluid coupling time Scale}

In the two-fluid model for superfluid $^4\text{He}$, the superfluid (density $\rho_\text{s}$, velocity $\bm{v}_\text{s}$) and normal fluid (density $\rho_\text{n}$, velocity $\bm{v}_\text{n}$) interact via mutual friction induced by quantized vortices. 
Focusing on the relaxation of the relative velocity $\bm{w} = \bm{v}_\text{n} - \bm{v}_\text{s}$ and neglecting advection, pressure gradients, and viscous diffusion, the simplified momentum equations with mutual friction force $\bm{F}_{\text{ns}}$ \cite{donnelly91} are
\begin{equation}
\rho_\text{n} \frac{\partial \bm{v}_\text{n}}{\partial t} = \bm{F}_{\text{ns}}, \quad \rho_\text{s} \frac{\partial \bm{v}_\text{s}}{\partial t} = - \bm{F}_{\text{ns}}.
\end{equation}
According to the Hall--Vinen model, the mutual friction force per unit volume is expressed as
\begin{equation}
\bm{F}_{\text{ns}} = \rho_\text{s} \kappa L \left[ B' \hat{\boldsymbol{\omega}} \times \bm{w} + B \hat{\boldsymbol{\omega}} \times (\hat{\boldsymbol{\omega}} \times \bm{w}) \right], \label{eq:mf_hall_vinen}
\end{equation}
where $\kappa$ is the quantum of circulation, $L$ is the vortex line density, $\hat{\boldsymbol{\omega}}$ is the unit vector along the vorticity, and $B, B'$ are the mutual friction parameters \cite{vinen57a,barenghi83}. In a turbulent state, isotropic averaging yields
\begin{equation}
\bm{F}_{\text{ns}} \approx - \frac{2}{3} \rho_\text{s} \kappa B L \bm{w}. \label{eq:mf_avg}
\end{equation}
The rate of change of relative velocity then becomes
\begin{equation}
\frac{\partial \bm{w}}{\partial t} = \bm{F}_{\text{ns}} \left( \frac{1}{\rho_\text{n}} + \frac{1}{\rho_\text{s}} \right) = -\frac{2}{3}\frac{\kappa B L \rho}{\rho_\text{n}} \bm{w},
\end{equation}
where $\rho = \rho_\text{s} + \rho_\text{n}$ is the total density. This equation expresses the exponential relaxation of relative velocity with the characteristic coupling time scale $\tau_{\text{coupling}}$ \cite{vinen00}:
\begin{equation}
\tau_{\text{coupling}} = \frac{\rho_\text{n}}{\rho \kappa B L} = \frac{3}{4 \alpha \kappa L}, \label{eq:tau_coupling}
\end{equation}
where $\alpha = B \rho_\text{n} / (2\rho)$ is the standard dimensionless mutual friction coefficient \cite{schwarz85}.

At high wavenumbers where viscous dissipation dominates the normal-fluid dynamics, the viscous time scale is given by $\tau_\nu(k) \sim 1/(\nu_n k^2)$, where $\nu_n$ is the kinematic viscosity of the normal fluid. The wavenumber $k_D$, beyond which mutual friction drives the high-wavenumber motion of the normal fluid through the drag from oscillating quantized vortices, is determined by the crossover condition $\tau_\nu(k_D) \sim \tau_{\text{coupling}}$:
\begin{equation}
\frac{1}{\nu_n k_D^2} \sim \frac{3}{ 4\alpha \kappa L} \quad \implies \quad k_D \sim \sqrt{\frac{4 \alpha \kappa L}{3\nu_n}}.
\end{equation}
Expressing $L$ in terms of the mean inter-vortex distance $\ell \sim L^{-1/2}$ (so that $k_\ell \sim 2\pi / \ell$), the crossover wavenumber scales as $k_D \sim \sqrt{\alpha \kappa / \nu_n} \, k_\ell$. At $T = 1.9\text{ K}$, substituting physical values ($\kappa \approx 1.0 \times 10^{-7}\text{ m}^2/\text{s}$, $\nu_n \approx 2.1 \times 10^{-8}\text{ m}^2/\text{s}$, and $\alpha \approx 0.21$) yields $k_D \approx k_\ell$.

For a localized normal-fluid vortex tube of volume $V = S D_z$ with cross-sectional area $S$ and height $D_z$ containing $N$ quantized vortex lines of length $D_z$, the bundled local vortex line density is $L_l = N/S$. The corresponding local coupling time scale governing the dynamics inside the vortex tube is given by
\begin{equation}
\tau_{\text{coupling},l} = \frac{3}{4 \alpha \kappa L_l}.
\end{equation}

\subsection{Supplemental movies}
Time evolution of quantized vortices (black lines) and normal-fluid vortex tubes (green isosurfaces for the second invariant of the velocity gradient tensor $Q = 1000~\mathrm{s^{-2}}$) from $t = 0$ to $20~\mathrm{s}$.

\noindent{\bf Movie S1:} Top view corresponding to Fig.1(a).

\noindent{\bf Movie S2:} Side view corresponding to Fig.1(b).

\bibliography{book,aps,other}

\end{document}